\documentclass[12pt,preprint]{aastex}

\usepackage{psfig}

\begin{document}

\newcommand{\msun}{M{$_{\odot}$}} 
\newcommand{\lsun}{L{$_{\odot}$}}
\newcommand{\rsun}{R{$_{\odot}$}}
\newcommand{\nduehp}{N$_2$H$^+$(1--0)}
\newcommand{\methduall}{CH$_3$OH(2$_{-1}$--1$_{-1}$)E2~\& (2$_0$--1$_0$)A$^+$}
\newcommand{\methcqall}{CH$_3$OH(5$_{-1}$--4$_{-1}$)E2~\& (5$_0$--4$_0$)A$^+$}
\newcommand{\methdu}{CH$_3$OH(2$_{k}$--1$_{k}$)}
\newcommand{\methcq}{CH$_3$OH(5$_{k}$--4$_{k}$)}
\newcommand{\lsim}{\;\lower.6ex\hbox{$\sim$}\kern-7.75pt\raise.65ex\hbox{$<$}\;}
\newcommand{\gsim}{\;\lower.6ex\hbox{$\sim$}\kern-7.75pt\raise.65ex\hbox{$>$}\;}

\slugcomment{Submitted to ApJ Letters}

\shorttitle{Near Infrared Spectral Classification of L-Dwarfs}
\shortauthors{Testi et al.}

\title{NICS--TNG low--resolution 0.85--2.45~$\mu$m spectra of L-Dwarfs:
a near-infrared spectral classification scheme for faint dwarfs}

\author{L. Testi\altaffilmark{1,6}, F. D'Antona\altaffilmark{2},
        F. Ghinassi\altaffilmark{3}, J. Licandro\altaffilmark{3},
	A. Magazz\`u\altaffilmark{3,4},
	R. Maiolino\altaffilmark{1}, F. Mannucci\altaffilmark{5},
	A. Marconi\altaffilmark{1}, N. Nagar\altaffilmark{1},
        A. Natta\altaffilmark{1} and E. Oliva\altaffilmark{1,3}
	}

\altaffiltext{1}{Osservatorio Astrofisico di Arcetri, Largo E.~Fermi 5,
I-50125 Firenze, Italy}
\altaffiltext{2}{Osservatorio Astronomico di Roma, via Frascati 33, I-00044,
Roma, Italy}
\altaffiltext{3}{Centro Galileo Galilei \& Telescopio Nazionale Galileo,
P.O. Box 565, E-38700, Santa Cruz de La Palma, Spain}
\altaffiltext{4}{Osservatorio Astrofisico di Catania, Via S. sofia 78, I-95123
Catania, Italy}
\altaffiltext{5}{CAISMI-CNR, Largo E. Fermi 5, I-50125 Firenze, Italy}
\altaffiltext{6}{ltesti@arcetri.astro.it}

\begin{abstract}
We present complete near-infrared (0.85--2.45~$\mu$m), low-resolution
($\sim$100) spectra of a sample of 26 disk L-dwarfs with reliable
optical spectral type classification. The observations have been
obtained with NICS at the TNG using a prism-based optical element
(the Amici device)
that provides a complete spectrum of the source on the detector.
Our observations show that
low-resolution near-infrared spectroscopy can be used to determine 
the spectral classification of L-dwarfs in a fast but accurate way.
We present a library of spectra that can be used as templates for
spectral classification of faint dwarfs. We also discuss
a set of near-infrared spectral indices well correlated
with the optical spectral types that can be used to accurately classify
L-dwarfs earlier than L6.
\end{abstract}


\section{Introduction}
\label{sintro}

The latest years have witnessed the discovery of numerous brown dwarfs close to
the Sun, in nearby clusters and associations, and in binaries. The strategy
of the optical and near-infrared imaging surveys (2MASS,  Kirkpatrick et
al.~\citeyear{Kea99}; \citeyear{Kea00}; the Sloan Digital Sky Survey, Fan et
al.~\citeyear{Fan00}; and DENIS, Delfosse et al.~\citeyear{Del97}, Tinney
et al.~\citeyear{Tin98}) has been so
successful that two new spectral classes (L and T) have been added to the
previous types, to help to classify very cool stellar objects.

For L-dwarfs, in spite of the remaining uncertainties in model atmospheres
for such cool objects (e.g. Leggett et al. 2001), it has been
possible to derive 
a detailed spectral
classification system in 9 subclasses 
from the systematic  changes observed in selected 
spectral features
(Kirkpatrick et al.~\citeyear{Kea99}, Mart\'{\i}n et al.~\citeyear{Mea99}) .
This spectral  classification has been developed in the red part
of the optical spectrum: the beginning of the L type is set by the weakening
of the TiO and VO bands, while the appearence of the CH$_4$\ bands signals
the transition to the T type.  However,
the optical spectral confirmation and classification of a candidate 
DENIS or 2MASS L-dwarf requires up to $\sim$1~hr of integration time with a
low-resolution
($\sim$1000) optical spectrograph at a large (10m-class) telescope,
depending on the spectral type and magnitude of the candidate. This  prevents
the applicability of the optical classification to deeper surveys.

Given that L-dwarfs emit most of their
radiation in the near-infrared bands from 1 to 2.5~$\mu$m, the
advantage of longer wavelengths is obvious. At present,
near-infrared spectra
are available  for a handful of objects (14) and very recently Reid et
al.~(\citeyear{Rea01}) attemped to establish  a near-infrared classification
scheme from full (0.9 to 2.5~$\mu$m) UKIRT spectra with resolution
$\sim$~500-1000; each spectrum required integration
times between 1 and 4~hours, depending on the spectral type of the star.
With a 4-m class telescope the time demand is comparable (or higher) to 
that required for the optical classification, and thus prohibitive for
large surveys. It is clear that intermediate- and high-resolution spectroscopy,
while necessary for investigating photospheric properties of selected
objects, is not suitable for candidate confirmation and classification 
of large, deep surveys.

In this paper we present low-resolution ($\sim$100)
near-infrared spectra of a sample of 26 L-dwarfs
with reliable optical spectral classification
from Kirkpatrick et al.~(\citeyear{Kea00}).
The spectra have been
obtained with a prism-based optical element (the Amici device), which provided
a complete near-infrared spectrum of each star in less than 15~min on source
at the italian Telescopio Nazionale Galileo (TNG), a 3.56-m telescope.
We describe the  observations in \S 2 and show in \S 3 that
low-resolution near-infrared spectroscopy can be used to determine 
the spectral classification of L-dwarfs in a fast but accurate way.
The potential of such an observing mode is discussed in \S 4, which
concludes the paper.

\section{Observations and results}
\label{sobs}

The observational data were collected at the 3.56m TNG
with the Near Infrared Camera and Spectrograph (NICS),
a cryogenic focal reducer designed as a near-infrared common-user instrument
for that telescope. The instrument is equipped with a Rockwell 1024$^2$
HAWAII near-infrared array detector. Among the many imaging and spectroscopic
observing modes (Baffa et al.~\citeyear{Bea00}), NICS offers a unique, 
high throughput, very low-resolution mode with an approximately constant
resolving power of $\sim$50, when the 1\arcsec\ wide slit is used. In this
mode a prism-based optical element, the Amici device, is used to obtain on
the detector a complete 0.85--2.45~$\mu$m long slit spectrum of the
astronomical source (Oliva~\citeyear{O01}). 


The 26 L-dwarfs in our sample 
cover in an approximately uniform way the 
optically defined spectral types ranging from L0 to L8.
All the selected sources are brighter than
K$_s\sim 14.4$, with 3 exceptions with K$_s$=14.5--14.8.
The sources were observed during the commissioning of
NICS in several observing runs from December~2000 to February~2001.
We used the 0\farcs5 wide slit and the resulting spectra have an
effective resolution of $\sim$100 across the entire spectral range.
Integration times on source varied from 4 to 15 minutes depending on the 
source brightness. Wavelength calibration was performed using an Argon lamp
and the deep telluric absorption features. The telluric absorption 
was then removed by dividing each of the object spectra by an A0
reference star spectrum observed at similar 
airmass, and normalized using a synthetic A0 star
spectrum smoothed to the appropriate resolution.
Four of the targets, which are also among the fainter in our
sample, were observed in unfavorable weather conditions resulting in a poor
compensation of the deep atmospheric features and noisier spectra.
The accuracy of the spectral shapes was checked by computing the expected 2MASS
colors from our spectra. When normalized at 
H band, our synthesized and the 2MASS fluxes at J and K$_s$ differ by less than 
2~$\sigma$ in all but three cases, where one of the two bands is more 
discrepant.


The final spectra are shown in
Figure~\ref{fspectra}. The objects are shown from top to bottom and from
left to right in order of increasing optical spectral type from L0 to L8.
The spectra have been normalized by the average flux in the 1.235--1.305~$\mu$m
region, a constant offset has been added to each one to avoid overlap.
The spectra show the same general features described in Leggett et
al.~(\citeyear{Lea01}) and Reid et al.~(\citeyear{Rea01}). In our low-resolution
spectra the atomic lines of Na~I and K~I and the FeH lines in the J-band are not
resolved, although their blended absorption features are clearly seen in the
early type dwarfs. The spectra are dominated by the H$_2$O features at
$\sim$0.95, $\sim$1.15, $\sim$1.40, $\sim$1.85, and $\sim$2.4~$\mu$m.
TiO, near 0.85~$\mu$m, and CO, longward of 2.3~$\mu$m, are visible in some
of the spectra, depending on spectral type and signal to noise. 

\section{The NIR classification scheme}
\label{sclass}

\begin{figure}
\epsscale{1.1}
\plotone{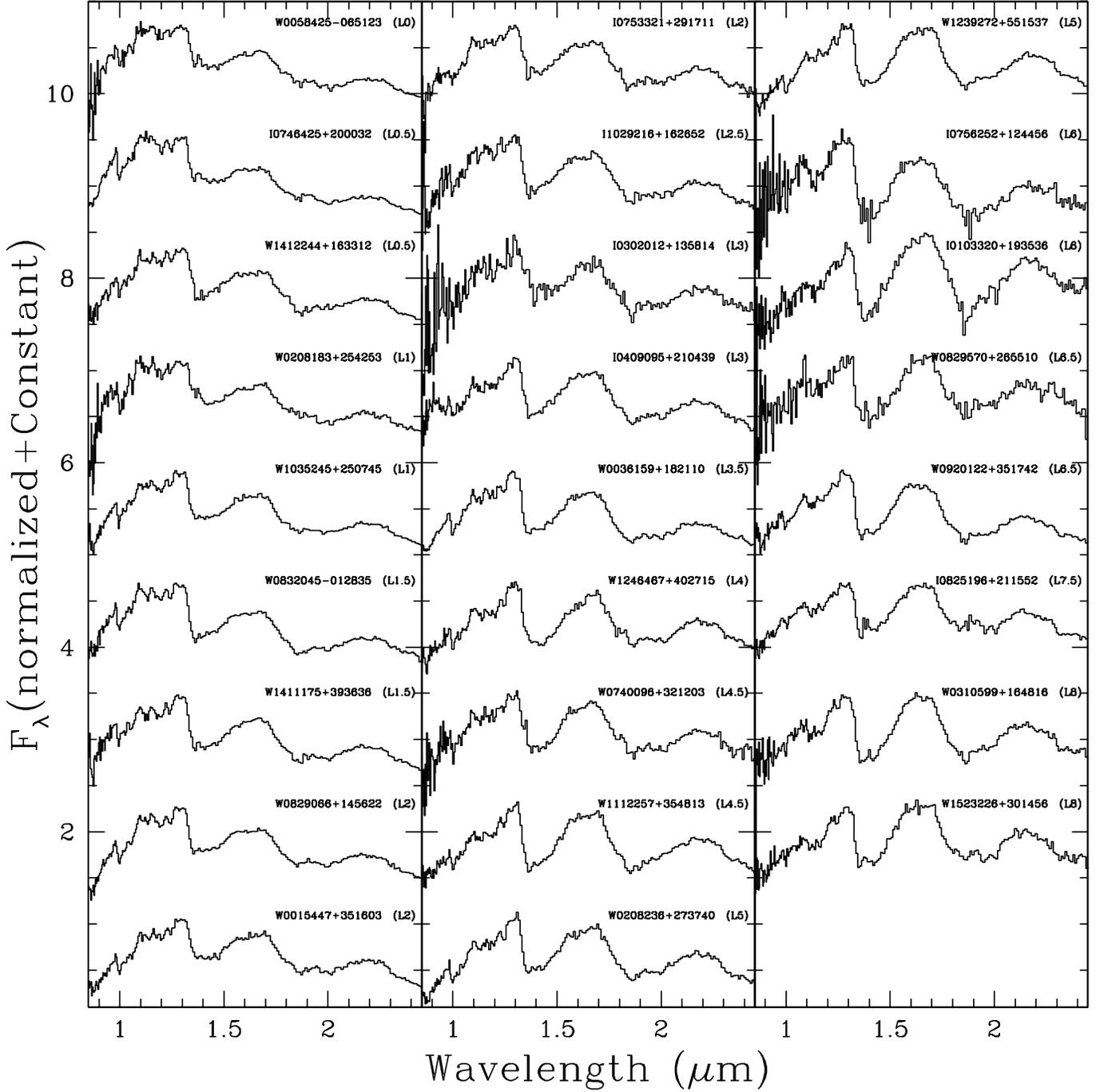}
\caption[f1.eps]{\label{fspectra}
0.85--2.45$\mu$m low-resolution near-infrared spectra for all the L-dwarfs in
our sample. All spectra have been normalized by the average flux 
between 1.235 and 1.305~$\mu$m and a
constant shift has been added to each to separate them vertically. Each
spectrum is labeled with the 2MASS name (the 2MASSJ prefix has been omitted)
and the optical spectral type from Kirkpatrick et al.~(\citeyear{Kea00}).
All spectra are available in electronic form upon request from the authors.
}
\end{figure}

\begin{figure}
\epsscale{0.57}
\plotone{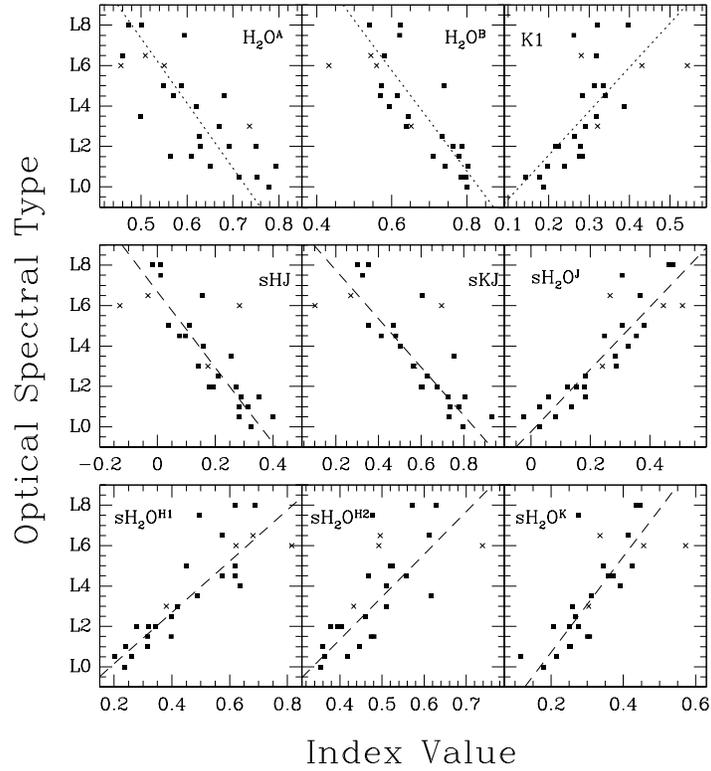}
\caption[f2.eps]{\label{fallidx}
Correlations between optical spectral types and the near-infrared
spectral indices. The top three panels show the H$_2$O$^A$, H$_2$O$^B$, and K1
indices calculated for the stars in our sample, the dotted lines show the 
linear fits of Reid et al.~(\citeyear{Rea01}). The bottom six panels show the
new indices defined in Table~\ref{tidx}; the dashed lines depict the linear
fits described in the text.  The four sources with poor telluric correction
spectra are shown as crosses.
}
\end{figure}

Despite their low-resolution, the spectra of Figure~\ref{fspectra} allow us
to identify a set of spectral indices that can be used to define a
near-infrared spectral classification scheme
which is well correlated with the widely used optical classification scheme of
Kirkpatrick et al.~(\citeyear{Kea99}) and Mart\'{\i}n et al.~(\citeyear{Mea99}).
A first attempt in this direction has already been taken by
Reid et al.~(\citeyear{Rea01}).
The main conclusion of 
their study is that while the J-band atomic lines are only weakly correlated
with the optical spectral types, it is possible to define indices based
on the H$_2$O wings which are well correlated with the optical types (at
least up to L6). 
For all the stars in our sample we computed the three indices that
Reid et al.~(\citeyear{Rea01}) found to be the most 
correlated with the optical spectral type:
K1 (see also Tokunaga \& Kobayashi~\citeyear{TK99}),
H$_2$O$^A$, and H$_2$O$^B$, all related to the strength
or slope of the water absorption features. In the top panels of
Fig.~\ref{fallidx}
we show the datapoints from our sample compared with the fits reported
by Reid et al.~(\citeyear{Rea01}); our spectra are generally consistent
with their fits.
Note that, as in Reid et al.~(\citeyear{Rea01}),
the K1 index can be used only for types earlier than L5; moreover
our data indicate saturation at late spectral types also for H$_2$O$^B$, while
H$_2$O$^A$ shows a very large scatter. It is possible that this behaviour 
of the H$_2$O$^A$ and H$_2$O$^B$ indices may be caused by the lower
resolution of our spectra.

\begin{figure}
\epsscale{0.45}
\plotone{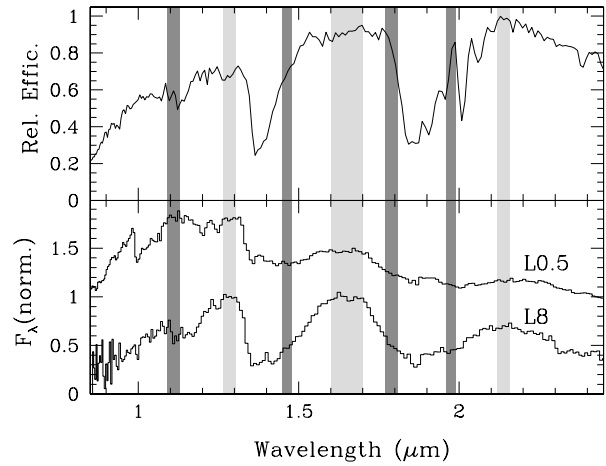}
\caption[f3.eps]{\label{fspidx}
Top panel: relative efficiency of the system, including atmosphere.
Bottom panel: spectra of I0746425$+$200032 (L0.5) and W0310599$+$164816 (L8).
Shaded in grey are the regions used to define the new indices of
Table~\ref{tidx},
dark grey regions correspond to the regions were the indices sample the water
wings. All indices avoid the worse telluric absorption regions.
}
\end{figure}

We also computed six additional 
indices which are best suited for low-resolution, complete spectra.
The new indices are defined in Table~\ref{tidx} in a similar way as
the Tokunaga \& Kobayashi~(\citeyear{TK99}) indices. Two of the new indices
(sHJ and sKJ) are based on the slope of the continuum, and can be reliably
defined using our spectra because the entire spectral range is observed 
simultaneously in the same atmospheric conditions, without the need of a 
problematic intercalibration of various spectral segments.
All the other indices measure the slope of the water line wings. They 
have been defined so as to avoid as much as possible the 
spectral regions affected by the worse telluric absorption.
To illustrate this point in Fig.~\ref{fspidx} we show the relative system efficiency
(including atmosphere), two of the spectra of Fig.~\ref{fspectra},
representative of the extreme classes (L0.5 and L8), and
the spectral regions used to define the various indices shaded in grey.
In Figure~\ref{fallidx} the
value of all the six indices are plotted against the optical spectral type
of each star. 
The sH$_2$O$^J$ index is a measure of the strength of the 
water absorption feature at 1.1~$\mu$m and, although it shows a very nice 
correlation with the optical spectral type in our data, it should be used 
with care as it may be seriously affected by a poor correction of the
telluric absorption. 
With only few exceptions, all stars with good
spectra show a tight correlation between the newly defined spectral 
indices and the optical spectral type. In Table~\ref{tidx} we also show the
linear relation between the optical spectral types and the index values;
the spectral types are coded as: L0$\equiv$1.0, L8$\equiv$1.8, with 0.1 
step per subclass. We did not attempt to fit again the indices of Reid et
al.~(\citeyear{Rea01}).
For the spectral range L0 to L6 the linear fits offer
a classification accurate to approximately half a subclass, at later spectral
types most of the water indices saturate and the classification based on the
fits is not as accurate and a direct comparison with the spectral library of
Fig.~\ref{fspectra} is preferable. 

It is interesting to note that the good correlation of our ``narrow-band''
continuum indices 
(sJH and sJK) with the spectral type desappears when
the broad-band colours are used (Kirkpatrick et al.~\citeyear{Kea00}).
This can be understood in terms of the competing effects of 
the reddening of the continuum at later spectral types and the increasing
absorption from water features, the latter mostly affects the H and
K$_s$ broad bands.


\section{Conclusions}
\label{sconc}

We have presented a library of 
complete 0.85--2.45~$\mu$m low-resolution ($\sim$100) spectra 
of 26 disk L-dwarfs. This kind of spectral library, and the spectral
indices we have defined, provide a unique  tool for the
identification and spectral classification of L-dwarfs from large, deep
surveys, where the number and magnitudes of potential candidates make other
techniques prohibitive, even at large telescopes.
As an example, we estimate that the VLT next-generation, low-resolution
near-infrared spectrograph (NIRMOS),
will allow one to measure in 1~hr the $0.9-1.7$~$\mu$m spectrum of 
faint (J$>21$) dwarfs, and classify them using one of our sHJ,
sH$_2$O$^J$ or sH$_2$O$^{H1}$ indices.

Finally, we want to stress the advantage of using a device that produces the
complete near-infrared spectrum in one shot, without the need for
intercalibration of
various spectral segments obtained in varying atmospheric conditions.
This makes it possible to measure
spectral indices that use the shape of the continuum, rather than the
lines. These continuum indices (sHJ and sKJ) turn out to be very sensitive
and reliable tools for the spectral classification of cool objects.
Our results suggest that a rough, but very efficient
spectral classification, could be obtained by narrow-band imaging through
filters corresponding to the flux ranges used in the definitions
of sHJ and sKJ in Table~\ref{tidx}.

\smallskip
\noindent
{\bf Acknowledgements:}  
This paper is based on observations made with the Italian Telescopio
Nazionale Galileo (TNG) operated on the island of La Palma by
the Centro Galileo Galilei of the CNAA (Consorzio Nazionale per
l'Astronomia e l'Astrofisica) at the Spanish Observatorio del
Roque de los Muchachos of the Instituto de Astrofisica de Canarias
Support from ASI grants ARS-99-15 and 1/R/27/00 to the Osservatorio di
Arcetri is gratefully acknowledged. It is a pleasure to thank the Arcetri and
TNG technical staff and the TNG operators for their assistance during the 
commissioning of NICS.

%
%
%
\begin{deluxetable}{lll}
\tabletypesize{\scriptsize}
\tablecaption{Spectral indices definitions and fits to optical spectral types.
\label{tidx}}
\tablewidth{0pt}
\tablehead{
\colhead{Index} & \colhead{Definition}   & \colhead{Fit to optical Sp. Type}
}
\startdata
sHJ&$\frac{<F_{1.265-1.305}>-<F_{1.60-1.70}>}
          {0.5(<F_{1.265-1.305}>+<F_{1.60-1.70}>)}$ & Sp=--1.87\,sHJ+1.67 \\
&\\
sKJ&$\frac{<F_{1.265-1.305}>-<F_{2.12-2.16}>}
          {0.5(<F_{1.265-1.305}>+<F_{2.12-2.16}>)}$ & Sp=--1.20\,sKJ+2.01 \\
&\\
sH$_2$O$^J$&$\frac{<F_{1.265-1.305}>-<F_{1.09-1.13}>}
   {0.5(<F_{1.265-1.305}>+<F_{1.09-1.13}>)}$ & Sp=+1.54\,sH$_2$O$^J$+0.98 \\
&\\
sH$_2$O$^{H1}$&$\frac{<F_{1.60-1.70}>-<F_{1.45-1.48}>}
   {0.5(<F_{1.60-1.70}>+<F_{1.45-1.48}>)}$ & Sp=+1.27\,sH$_2$O$^{H1}$+0.76 \\
&\\
sH$_2$O$^{H2}$&$\frac{<F_{1.60-1.70}>-<F_{1.77-1.81}>}
   {0.5(<F_{1.60-1.70}>+<F_{1.77-1.81}>)}$ & Sp=+2.11\,sH$_2$O$^{H2}$+0.29 \\
&\\
sH$_2$O$^K$&$\frac{<F_{2.12-2.16}>-<F_{1.96-1.99}>}
   {0.5(<F_{2.12-2.16}>+<F_{1.96-1.99}>)}$ & Sp=+2.36\,sH$_2$O$^{K}$+0.60 \\
\enddata
\tablecomments{$<F_{\lambda_i-\lambda_j}>$ is the average flux in the range 
$\lambda_i$ to $\lambda_j$.}
\end{deluxetable}

%
\clearpage

\clearpage

\clearpage

\end{document}